\documentclass[%
 reprint,
superscriptaddress,
amsmath,amssymb,
aps,
prb,
floatfix
]{revtex4-2}

\usepackage{multirow}
\usepackage{graphicx}
\usepackage{dcolumn}
\usepackage{bm}
\usepackage[version=3]{mhchem}
\usepackage{nicefrac}

\usepackage{color}
\usepackage{dcolumn}
\usepackage{amssymb}
\usepackage{url}
\usepackage[colorlinks=True,linkcolor=red,citecolor=blue,urlcolor=blue]{hyperref}
\usepackage{xfrac}
\usepackage{tabularx}
\usepackage{xspace}
\usepackage{amsmath}
\usepackage[normalem]{ulem}
\usepackage{mathtools}
\usepackage{float}
\usepackage{booktabs}
\usepackage{physics}
\usepackage{xspace}
\newcommand{\cuag}{\ce{CuAg(SO4)2}\xspace}

\usepackage{pdfpages}
\makeatletter
\AtBeginDocument{\let\LS@rot\@undefined}
\makeatother

\begin{document}

\title{Highly unusual, doubly-strongly-correlated, altermagnetic, 3D analogue of parent compounds of high-$T_{\rm c}$ cuprates}
\author{Harald O. Jeschke}
\affiliation{Research Institute for Interdisciplinary Science, Okayama University, Okayama 700-8530, Japan}

\author{Makoto Shimizu}
\affiliation{Department of Physics, Graduate School of Science, Kyoto University, Kyoto 606-8502, Japan}

\author{Igor I. Mazin}
\affiliation{Department of Physics \& Astronomy, George Mason University, Fairfax, VA 22030, USA and
Quantum Science and Engineering Center, George Mason University, Fairfax, VA 22030, USA.}
\date{\today}

\begin{abstract}
Discovery of high-temperature superconductivity (HTSC) in strongly correlated cuprates opened a new chapter in condensed matter physics, breaking existing stereotypes of what is a material base for a good superconductor (``Matthias rules''), at the same time emphasizing richness and challenge of strongly correlated physics, personified by the most strongly correlated $3d$ ion, Cu$^{2+}$. A recently reported new compound, CuAg(SO$_4$)$_2$, combines in a fascinating way the same ion with the most strongly correlated $4d$ one, Ag$^{2+}$. In this Letter, we present a detailed analysis of electronic and magnetic properties of this material, and show that it is very different from the HTSC cuprates in several different ways, and opens a door into further research of superconductivity and magnetism, in particular altermagnetism, in strongly correlated materials.
\end{abstract}

\maketitle

{\it Introduction.-}
Four decades ago the world was tantalized by the discovery of the high-critical-temperature superconductors. It was soon appreciated that a pivotal role in the physics of these materials was played by the Cu$^{2+}$ ion in a $3d^9$ configuration, a strongly-correlated spin-1/2 object with one rather localized hole in the Cu $3d_{x^2-y^2}$ orbital, and that the magnetic interaction between these ions, generated by the oxygen-mediated superexchange processes and peaked in the 2D momentum space at $\mathbf{q}=(\pi,\pi)$, is instrumental in understanding its properties~\cite{Dagotto1994,Imada1998}. 

Initial microscopic theories of these materials, rather abundant, relied upon a simple single-band Hubbard Hamiltonian, with a Mott insulator as a parent compound~\cite{Lee2006}. However, it was then realized that, while close to Mott insulators, the parent compounds belonged to a different class, namely charge-transfer insulators (CTI)~\cite{Zaanen1985}. Indeed, the top of the O-$p$ band appeared above the lower Hubbard band (LHB), albeit not by much, which led to important ramifications.

The Cu$^{2+}$ valence state occurs in many natural minerals as well as in synthesized compounds. 
Nevertheless, the peculiar physics associated with the $d^9$ band occupancy inspired vigorous searches for other $d^9$ materials. So far, the lion share of this activity was associated with Ni$^+$ compounds, also having a $d^9$ configuration, albeit less localized than in Cu$^{2+}$~\cite{Hansmann2009,Li2019}. At the same time, solid state chemists had their eyes on a heavier analogue of  Cu$^{2+}$, namely  Ag$^{2+}$~\cite{Gawraczynski2019}. The $d$-hole in this state also highly localized, and materials with Ag$^{2+}$ are truly rare.

Thus, the recent experimental report of a new $d^9$ compound forming a new, fourth class (counting Cu$^{2+}$, Ni$^+$ and Ag$^{2+}$ as the first three), {\cuag}~\cite{Domanski2023}, opens an exciting opportunity of a new variation on the old theme: a combination of  Cu$^{2+}$ {\it and} Ag$^{2+}$ in the same compound warrants close attention. Moreover, as we discuss later in the paper, magnetic order in this material belongs to a recently discovered class of altermagnets\cite{Mazin2022,Smejkal2022}, adding an additional dimension of interest to this material. It is worth noting that the only altermagnet in this class discussed so far is \ce{La2CuO4}, where altermagnetism appears only because of small rotations of the CuO$_6$ octahedra. In contrast, in {\cuag}, as discussed below, altermagnetism appears already in the Cu-Ag sublattice.

\begin{figure*}[hbt]
    \includegraphics[width=0.95\textwidth]{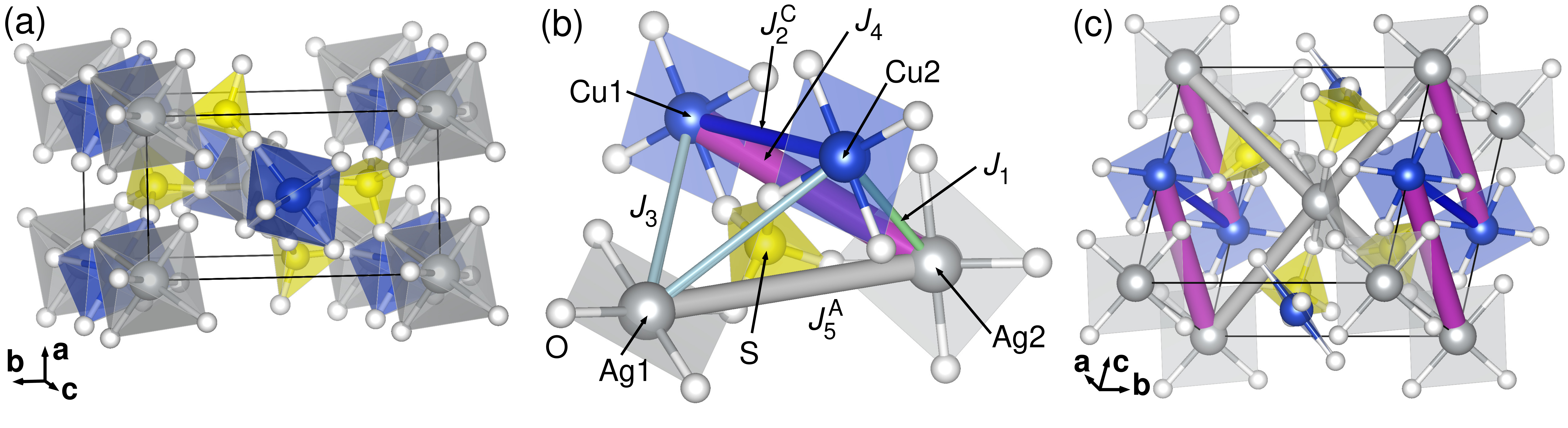}
    \caption{(a) Crystal structure of \cuag, as reported in Ref.~\cite{Domanski2023}. 
    (b) Structural unit showing only one SO$_4$ cluster with its nearest neighbor connected via $d_{x^2-y^2}$ orbitals. Cu and Ag sites are only numbered for use in Table~\ref{tab:measures}.
    (c) Illustration of relevant exchange paths in the structure of {\cuag}.} 
    \label{fig:str}
\end{figure*}

One can summarize (Table \ref{tab:0}) the key differences distinguishing {\cuag} from cuprate superconductors, as demonstrated and discussed in detail in this Letter:
\begin{table}[htb]
    \centering
    \begin{tabular}{l|c|c}
    \hline
         & parent cuprates  & \cuag \\
         \hline
    strongly correlated species     & one (Cu) & two (Cu, Ag)\\
    excitation gap   & intermediate,& strongly CT\\
       & closer to CT& \\
    leading superexchange     & Cu--O--Cu & $M$--SO$_4$--$M$\\
     path  & &  \\
    leading superexchange   & first & $3^{\rm rd}$, $5^{\rm th}$ and $6^{\rm th}$ \\
     neighbors   & &  \\
     leading   superexchange & $\sim $2.7--2.8\,\AA & $5.7, 6.0, 4.7$\,\AA \\
     length$^*$   & &  \\
     dimensionality  & 2D & 3D\\
          leading  spin fluctuations  & ${\bf q}=(\pi,\pi)$ & ${\bf q}=(0,0,2\pi)^\S$\\
      altermagnetism   & sometimes$^\dagger$ & yes$^\ddagger$ \\
\hline    
\end{tabular}
    \caption{
    Comparison between parent materials of cuprate superconductors and \cuag.\\
    \rule{\linewidth}{0.4pt}
    $^*$ in order of decreasing strength; $^\S$ in the extended Brillouin zone, corresponding to the intracell magnetic order; $^\dagger$ in \ce{La2CuO4} and similar materials, due to O octahedra rotations; $^\ddagger$ regardless of the presence of ligands.}
    \label{tab:0}
\end{table}

Given such unique properties of this just recently discovered compound, one should expect more experimental work in the nearest future. The goal of this Letter is to guide and inform these researches about fundamental electronic and magnetic properties of this material. In the next Section we will present and discuss its electronic structure, then we will turn to magnetic interactions in the system, and demonstrate and explain their highly unusual topology. We will then identify the stable ground state magnetic configuration and discuss its properties, including the character of spin fluctuations once the static order is suppressed (e.g., by doping). 

{\it Crystal structure.-}
The crystal structure is formed by chains aligned along the $c$ axis of octahedral-coordinated Cu and Ag, with edge-sharing octahedra (as opposed to layered perovskite cuprates where octahedra are corner-sharing), Fig.~\ref{fig:str}\,(a). 
These chains are bridged by SO$_4$ radicals, forming ``molecular ligands'', which are nearly ideal tetrahedra with S nearly central. The intra-, as well as interchain hoppings proceeds via these tetrahedra. \ce{CuO6} and \ce{AgO6} octahedra are strongly elongated (20\% for Cu, 30\% for Ag, cf. 28\% in La$_2$CuO$_4$), so that the $d$-holes reside in a well-defined
$d_{x^2-y^2}$ state.
Fig.~\ref{fig:str}\,(b) shows the minimal connectivity cluster, that is, an individual SO$_4$ 
tetrahedron with four metals attached to it.  Interestingly, all four metal ions are positioned geometrically different, as Table \ref{tab:measures} illustrates, and form different bond angles. 
\begin{table}[h!t]
    \centering
    \begin{minipage}{0.3\linewidth}
        \begin{tabular}{c|r}
        \toprule
         $M$ & $M$-O-S \\
         \hline
         Cu1 & 135 \\
         Cu2 & 137 \\
         Ag1 & 123 \\
         Ag2 & 140 \\
         \hline
        \bottomrule
        \end{tabular}
    \end{minipage}
    \begin{minipage}{0.6\linewidth}
        \begin{tabular}{c|rrr}
        \toprule
             & $\theta$\,(deg.)  & $d$\,({\AA}) & $J$\,(K) \\
             \hline
             Cu1-S-Cu2 & 96 & 4.73 & 34 \\
             Cu1-S-Ag1 & 102 & 4.96 & 5 \\
             Cu1-S-Ag2 & 122 & 5.73 & 166 \\
             Cu2-S-Ag1 & 102 & 4.96 & 5 \\
             Cu2-S-Ag2 & 66 & 3.58 & -3 \\
             Ag1-S-Ag2 & 134 & 6.02 & 92 \\
             \hline
        \bottomrule
        \end{tabular}
    \end{minipage}
    \caption{
      Left: angles formed by the $M$--O--S in degrees.
      Right: angles $\theta$ in degrees, distance $d$ in {\AA} of corresponding path and calculated (see the section of Magnetic interactions for details) exchange coupling constants $J$ in K formed by $M$--S--$M$.
    }
    \label{tab:measures}
\end{table}

\begin{figure}[hb]
    \includegraphics[width=0.95\columnwidth]{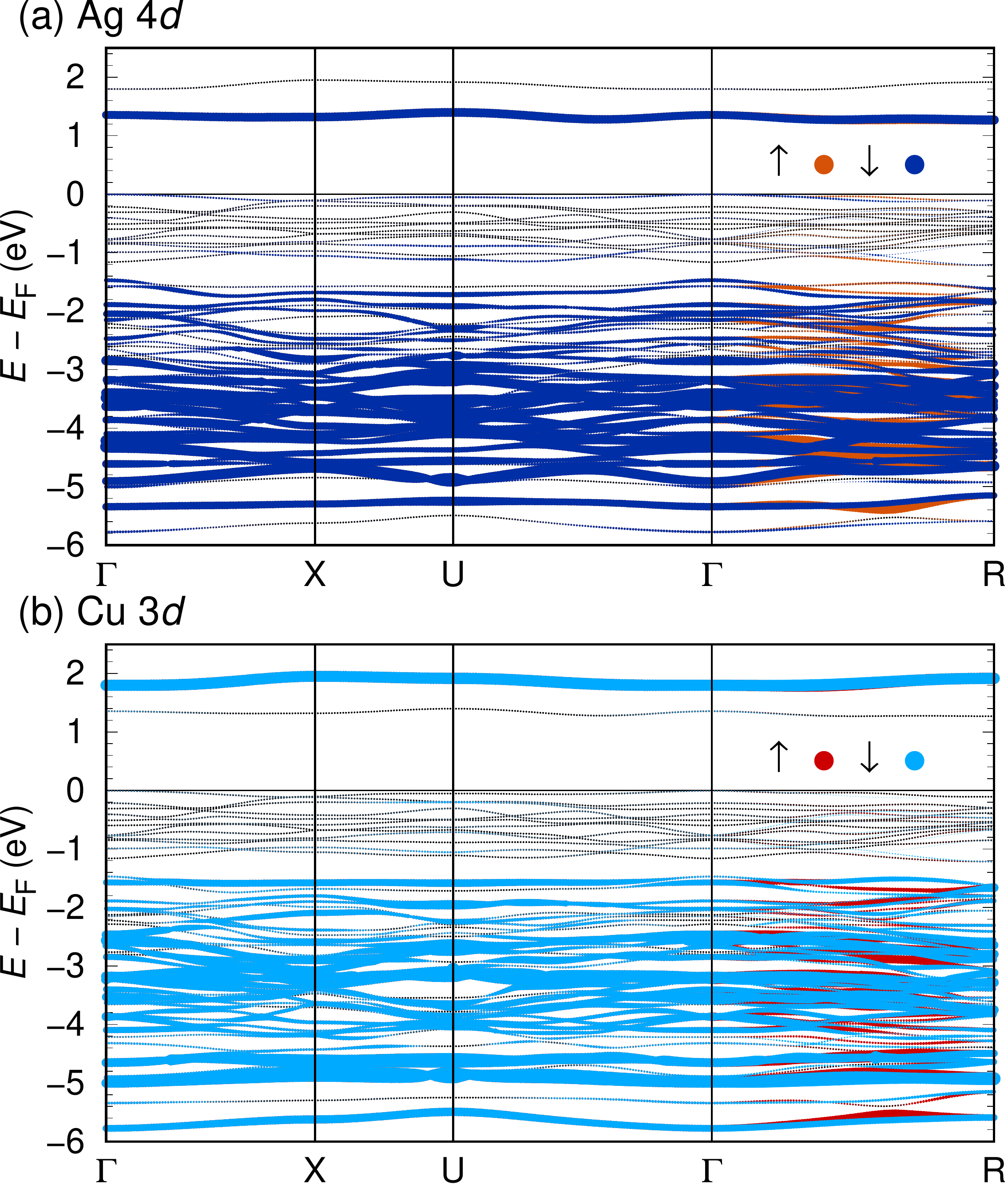}
    \caption{GGA+U band structure of {\cuag} in the lowest energy AFM state. Colors red (spin $\uparrow$) and blue (spin $\downarrow$) indicate (a) Ag $4d$ orbital weights and (b) Cu $3d$ orbital weights. The altermagnetic property of {\cuag} is clear from the spin splitting along $\Gamma$--$R$ path.}
    \label{fig:bands}
\end{figure}
{\it Electronic structure.-}
The calculated band structure is shown in Fig.~\ref{fig:bands}, and the corresponding density of states (DOS) in Fig.~\ref{fig:dos}. For comparison, the DOS for La$_2$CuO$_4$ (calculated with the same setup) are shown in the Supplementary information). 
Several interesting features manifest themselves. First, due to much longer hopping paths, and strong covalent bonding in the SO$_4$ cluster, O bands are pushed up, compared to HTSC cuprates, and are twice (!) narrower. As a result, twelve O $p$ bands are separated from the rest by a full gap, and are much more pure O $p$ than in the cuprates, while the charge-transfer (CT) gap is much larger (1.3 eV vs. 0.4 eV, for the same parameter choice) and the upper Hubbard bands much narrower in {\cuag}. As a result, the metal states are more correlated, and the CT character more pronounced than in the cuprates, promising interesting ramifications.

These new features can also be traced down to the fact that the actual ``ligand'' in this system is in fact the sulfate radical, which has interesting molecular orbital structure\cite{Bishop1967,HOJER}: one triple-degenerate $t_1$ orbital in each spin, which is pure O $p$ by symmetry, and also a mixed O-S one, also a triplet, $3t_2$. The latter is the higher occupied orbital {\it if S d is not included}\cite{Bishop1967}. However, the high-lying S $d$ pushes this state down\cite{Bishop1967}, resulting in clear separation of the upper half of the O bands, well above the metal $d$ bands, and the lower half, overlapping the latter. What is important here is that if the system is doped by holes, they will be purely O $p$, as opposed to cuprates, where they are considerably mixed with Cu d. 

One other observation from Fig. \ref{fig:bands} is that the $d$-bands along the $\Gamma$-R line are spin split, despite the material being antiferromagnetic, and centrosymmetric. Indeed, one can observe that the symmetry operation that maps the spin up and spin down sublattices in the P2$_1/c$ group is the $c$ glide, while inversion maps each spin upon itself. This is the condition for altermagnetism\cite{checker}, a new phenomenon actively discussed in recent months. It is worth noting that some of the cuprates, most notably La$_2$CuO$_4$ are also altermagnets, but there this feature appears of the result of the CuO$_6$ octahedra rotations, and disappears when the structure becomes tetragonal under doping. Here, however, it is robust and present even if ligands are entirely removed (the Cu-Ag sublattice still has the same symmetry).

\begin{figure}[h]
    \includegraphics[width=0.95\columnwidth]{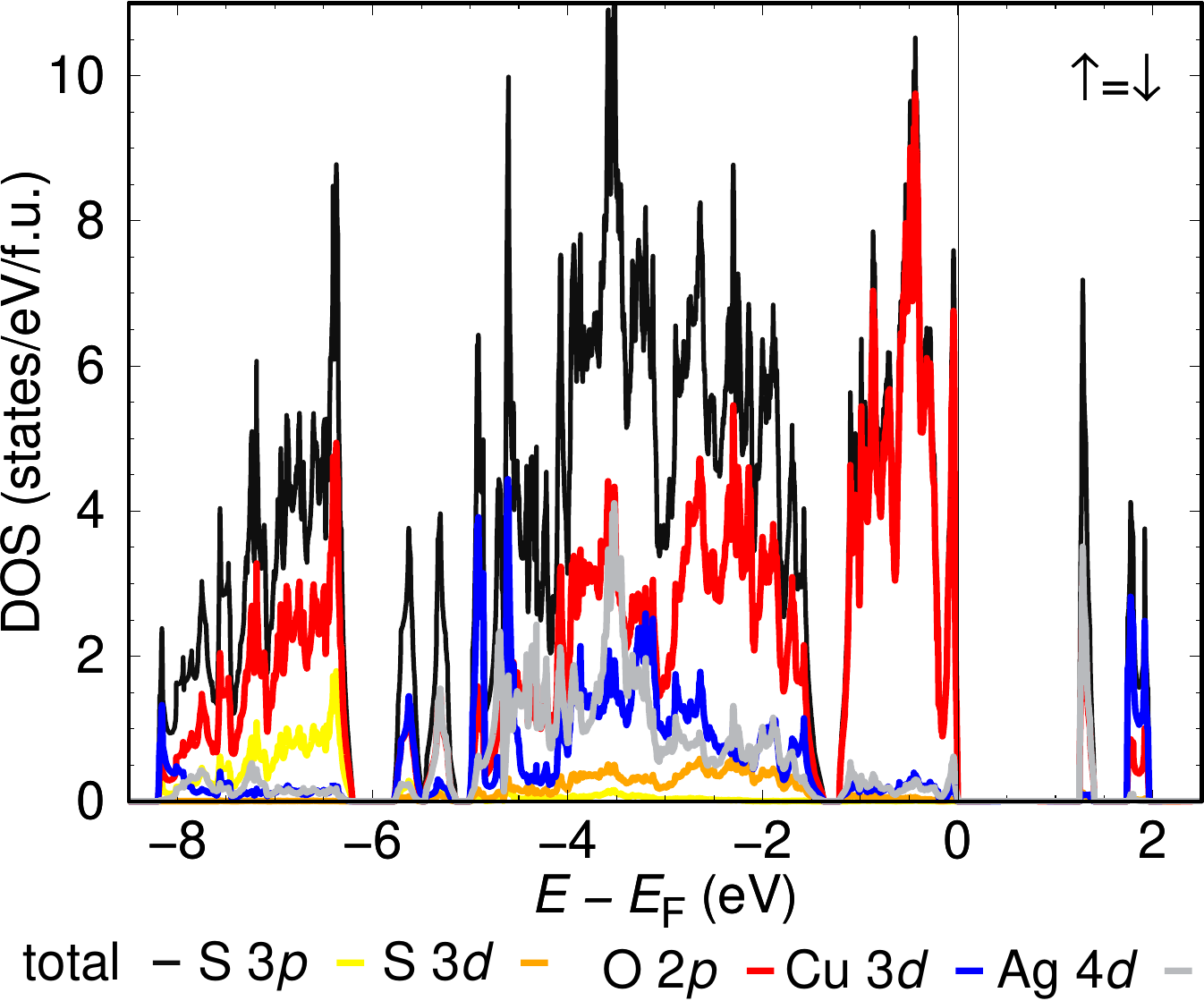}
    \caption{GGA+U density of states per spin of {\cuag} in the lowest energy AFM state. Spin $\uparrow$ and spin $\downarrow$ are identical so only spin $\uparrow$ is shown.  }
    \label{fig:dos}
\end{figure}

\begin{table*}[ht]
    \begin{tabular}{c|c|cc|c|c|cc|c|cc|cc|c|c|cc}
    \toprule
    name & $J_1$ & $J_2^{\rm A}$ & $J_2^{\rm C}$ & $J_3$  & $J_4$ & $J_5^{\rm A}$ & $J_5^{\rm C}$ & $J_6$ & $J_7^{\rm A}$ & $J_7^{\rm C}$ & $J_8^{\rm A}$ & $J_8^{\rm C}$ & $J_{10}$ & $J_{13}$ & $J_{16}^{\rm A}$ & $J_{16}^{\rm C}$ \\
 $M1\,M2$   & AgCu&AgAg &CuCu &AgCu &AgCu  &AgAg &CuCu &AgCu &AgAg&CuCu&AgAg&CuCu&AgCu&AgCu&AgAg&CuCu\\
       $d_{M1\,M2}$\,(\AA) &3.579 &\multicolumn{2}{|c|}{4.734} &4.961&5.727 & \multicolumn{2}{|c|}{6.017}&6.134&\multicolumn{2}{|c|}{6.215}&\multicolumn{2}{|c|}{7.158}&8.332&8.846&\multicolumn{2}{|c}{9.266}\\
        \hline
$J$ (K)& -3 & 3&34  &5 &168 &92 &-1 &-3&0&-4&-2&0&0&0&1&0\\
    \bottomrule
    \end{tabular}
    \caption{Exchange parameters for {\cuag} determined by DFT based energy mapping. The on-site interaction values $U_{Ag}=3.65$\,eV and $U_{\rm Cu}=4.87$\,eV are chosen in order to match the experimental value of the Curie-Weiss temperature of $\theta_{\rm CW}=-140$\,K~\cite{Domanski2023}. The $U_{\rm Cu}$ is smaller than typical values for $Cu^{2+}$ which are often in the range $6\,{\rm eV}\le U \le 8\,{\rm eV}$; this occasionally happens when $U$ is chosen to match a $\theta_{\rm CW}$ energy scale in energy mapping. }
    \label{tab:exchangepars}
\end{table*}

\begin{figure} [h!t]
    \includegraphics[width=\columnwidth]{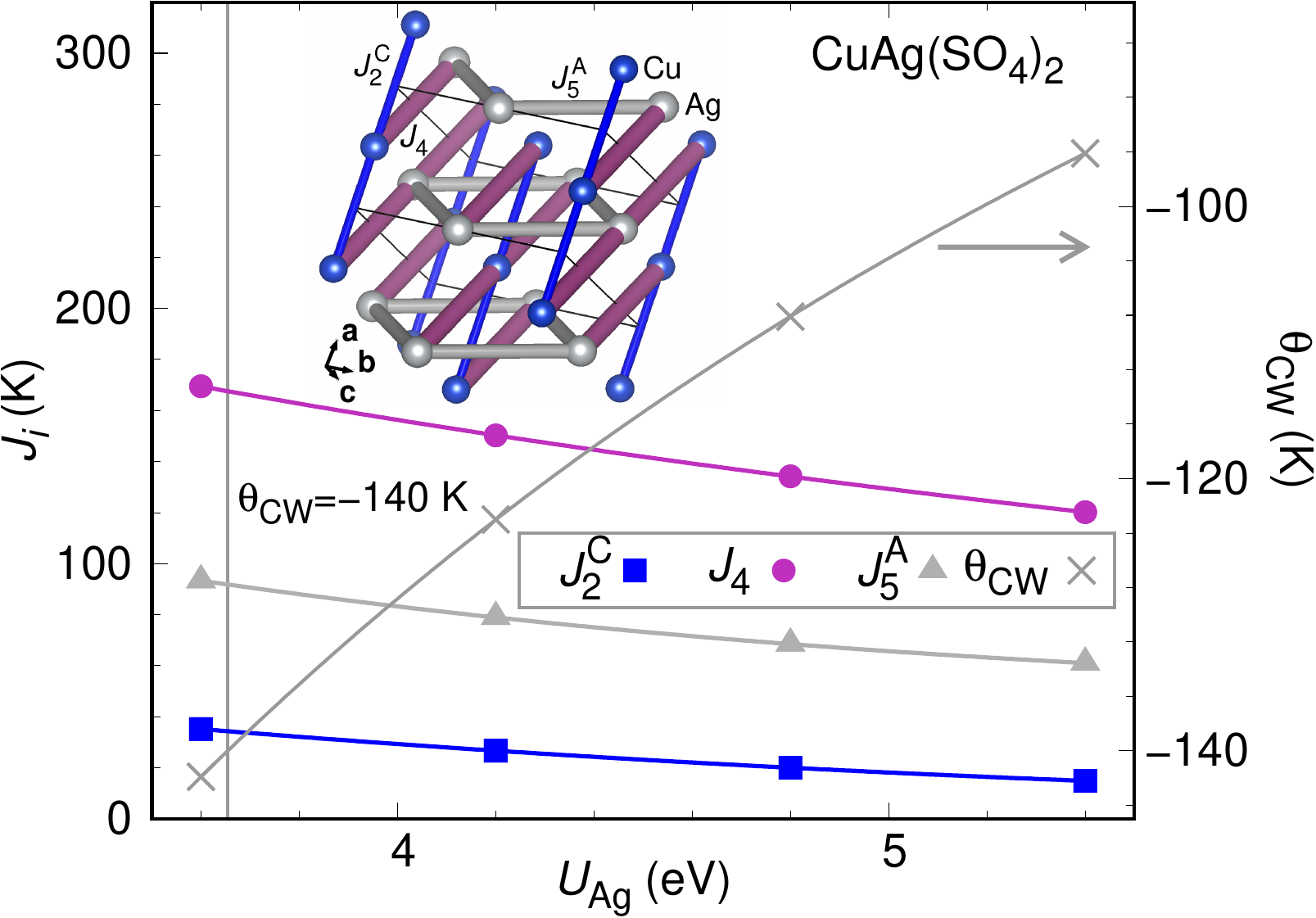}
    \caption{DFT energy mapping result for {\cuag}. Most important exchange interactions for four different values of on-site interaction $U_{\rm Ag}$ at fixed Hund's rule coupling strengths $J_{\rm H}=1$~eV for Cu and $J_{\rm H}=0.75$~eV for Ag. $U_{\rm Ag}$ is fixed at 75\% of $U_{\rm Cu}$. Inset: Exchange paths of {\cuag} as defined by the three dominant exchange interactions. Width of bonds is chosen so that cross section is proportional to the strength of the coupling.  
    }
    \label{fig:couplings}
\end{figure}

{\it Magnetic interactions.-}
We determine parameters of the Heisenberg Hamiltonian
  $  H=\sum_{i<j} J_{ij} {\bf S}_i\cdot {\bf S}_j$
for {\cuag} using DFT energy mapping. This approach has provided very good results for many Cu$^{2+}$ $S=1/2$ magnets~\cite{Chillal2020,Heinze2021,Hering2022}, so we can expect it to work for {\cuag} as well. We make sure to capture all relevant exchange interactions by resolving all couplings up to twice the nearest neighbor Cu-Ag distance. For this purpose, we use a 5-fold supercell containing ten formula units. For the DFT+U functional, we need onsite interactions and values of the Hund's rule coupling for both Cu$^{2+}$ and Ag$^{2+}$. Between Cu $3d$ orbitals and Ag $4d$ orbitals, we introduce a factor 0.75 which is reasonable to account for the better screening in the heavier ion. For Cu$^{2+}$, we use the typical value $J_{\rm H}^{\rm Cu}=1$\,eV that has yielded good agreement with experiment in many cases. Fig.~\ref{fig:couplings}\,(a) shows the result of the energy mapping for four values of $U$.

Other exchange interactions besides the three we show are 3{\%} of the dominant coupling $J_4$ or less (\ref{tab:exchangepars}). We select values of $U_{\rm Ag}=3.65$\,eV, $U_{\rm Cu}=4.87$\,eV by demanding that the the full set of couplings matches the experimentally determined Curie-Weiss temperature of {\cuag} which is $\theta_{\rm CW}=-140$\,K~\cite{Domanski2023}. Note that these $U$ values shoukd be viewed as internal LDA+U parameters and not as spectroscopic $U$ values; they would be chosen differently if future experiments lead to a revision of the $\theta_{\rm CW}$ value. The inset of Fig.~\ref{fig:couplings} illustrates the lattice defined by $J_4$, $J_5^{\rm A}=0.55J_4$ and $J_2^{\rm C}=0.20J_4$. The Hamiltonian is dominated by antiferromagnetic Cu-Ag chains (purple) which are linked by AFM Ag-Ag square lattices. These two couplings can be satisfied by an AFM state where both Cu and Ag sublattices are AFM. However, the third strongest (but considerably smaller) coupling, an AFM Cu-Cu exchange, is moderately frustrating this Hamiltonian.

This results seems, on the first glance, counterintuitive. The strongest coupling comes from the fifth neighbors, and the two shortest bonds contribute practically nothing. To understand this we recall that the active orbitals here are $x^2-y^2$, and replot Fig.~\ref{fig:str}\,(a) using instead of the metal-centered octahedra only the squares corresponding to these orbitals (Fig. \ref{fig:str}\,(b)). One can see that these orbitals do not overlap on any oxygen, thus not generating any $M$--O--$M$ superexchange, but only via SO$_4^{2-}$ ions. This yields five longer-range superexchange paths, which include the three leading ones, plus two more that appear to be numerically small due to accidental cancellation of various hopping processes. 
As discussed in the previous section, electronically this material is in a strong charge transfer regime, so that instead of the standard Anderson's superexchange proportional to $t^4/(E_d-E_p)^2U$, where $E_d-E_p\gg U$, and $t$ is the characteristic metal-ligand hopping one gets~\cite{ginyat} $t^4/\Delta^3$, with $\Delta\ll U$ (note that the charge transfer energy $\Delta$ is smaller because the highest occupied level in sulfate is higher than in oxygen). Therefore, despite a relatively small effective $M$--S hopping the resulting interaction is sizeable.

Given the nontrivial exchange Hamiltonian, it is instructive to re-derive the classical molecular field (Weiss) theory specifically for this case. As usual, we introduce the Curie susceptibility $\chi(T)=\mu_{\rm eff}^{2}/3T,$ where $\mu_{\rm eff}
^{2}=3$ for $S=1/2$. Let us assume that in an external field $H$ the two sublattices, Ag and Cu,
acquire magnetic moments $M_\mathrm{A}$ and $M_\mathrm{C}.$ From the previous section, in
order of decreasing magnitude, the relevant exchange constants are $J_\mathrm{AC}=J_{4},$ $J_\mathrm{AA}=J_{5}^\mathrm{A}$
and $J_\mathrm{CC}=J_{2}^\mathrm{C}.$ The molecular field on the site A will be $2M_\mathrm{C}%
J_{4}+4M_\mathrm{A}J_{5}^\mathrm{A},$ on C  $2M_\mathrm{A}J_{4}+2M_\mathrm{C}J_{2}^\mathrm{C}.$ Thus the Weiss equation
will be%
\begin{align}
M_\mathrm{A}  & =(H+2M_\mathrm{C}J_{4}+4M_\mathrm{A}J_{5}^\mathrm{A})\chi\\
M_\mathrm{C}  & =(H+2M_\mathrm{A}J_{4}+2M_\mathrm{C}J_{2}^\mathrm{C})\chi
\end{align}

Solving for $M,$ the Curie-Weiss susceptibility
\begin{align}
\chi_\mathrm{CW}  & =(M_\mathrm{A}+M_\mathrm{C})/2H\nonumber\\
& =\chi\frac{\chi^{-1}-2J_{4}+J_{5}^\mathrm{A}+J_{2}^\mathrm{C}}{\chi^{-1}+2J_{5}^\mathrm{A}+J_{2}^\mathrm{C}+2\chi
(2J_{5}^\mathrm{A}J_{2}^\mathrm{C}-J_{4}^{2})}\label{1}
\end{align}
Expanding $1/\chi_\mathrm{CW}$ in $1/T$, we get the  Curie-Weiss law with the same $\mu_{\rm eff}=\sqrt{3}$
and  $\theta_\mathrm{CW}=(2J_{4}+2J_{5}^\mathrm{A}+J_{2}^\mathrm{C})S(S+1)/3=(2J_{4}+2J_{5}^\mathrm{A}+J_{2}^\mathrm{C})/4$, the expression that we used above to fix $U_{\rm eff}$. The reason why Ref. \cite{Domanski2023} found a surprisingly large $\mu_{\rm eff}^{2}%
=2.3^{2}=5.3,$ corresponding to $S=0.75,$ rather than $S=1/2$, is unclear at this point.

\begin{figure*}[htb]
\includegraphics[width=\linewidth]{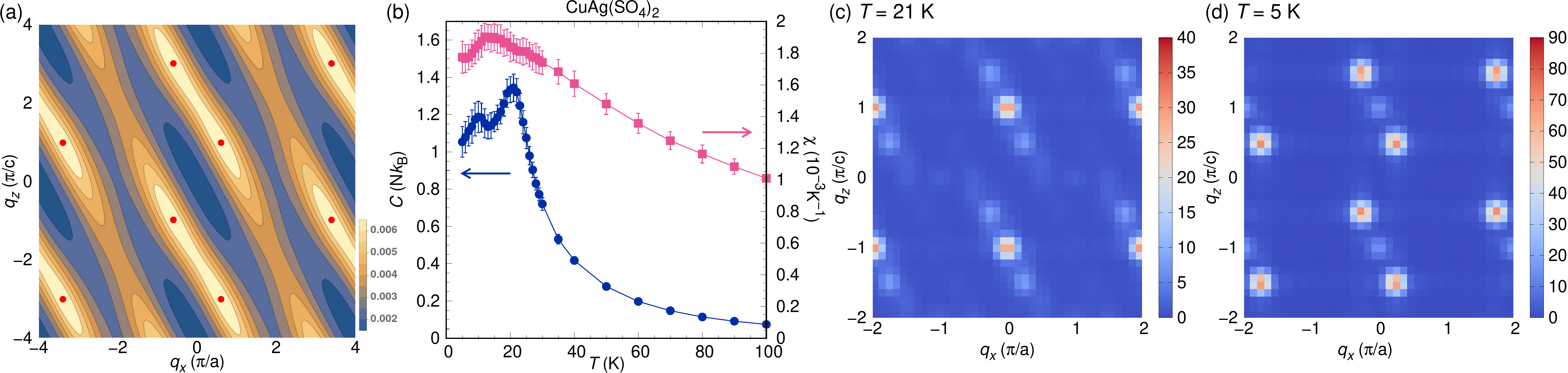}
    \caption{Susceptibility and classical Monte Carlo results for {\cuag}. (a) Susceptibility $\chi({\bf q},T)$ of {\cuag} at $T=300$\,K estimated using Eq.~\ref{eq:chiq}. (b) CMC specific heat and susceptibility. (c) and (d) correspond to spin structure factors at $T=21$\,K and $T=5$\,K, respectively. Note that for the ${\bf q}$ vectors in (a), (c) and (d) we neglect the small monoclinic angle of 94$^\circ$ and the difference in $a$ and $c$ lattice parameters. }
    \label{fig:CMC}
\end{figure*}

{\it Susceptibility.-}
We analyze the Hamiltonian by defining a strong coupling susceptibility~\cite{Otsuki2019} as $\chi({\bf q},T)=1/[T+J({\bf q})]$, {where}
\begin{equation}\begin{split}
  J({\bf q})&=2 J_1 \cos\frac{q_z}{2} + (J_2^\mathrm{A} + J_2^\mathrm{C}) \cos q_x\\&+  \bigg(2 J_3 \cos\frac{q_x}{2} + (J_7^\mathrm{A} + J_7^\mathrm{C}) \cos\frac{q_x-q_z}{2} \\&+ (J_5^\mathrm{A} + J_5^\mathrm{C}) \cos\frac{q_x+q_z}{2}\bigg) 2 \cos\frac{q_y}{2}\\&+  2 J_4 \cos\big(q_x+\frac{q_z}{2}\big) + 2 J_6 \cos\big(q_x-\frac{q_z}{2}\big) \\&+  (J_8^\mathrm{A} + J_8^\mathrm{C}) \cos q_z 
 \label{eq:chiq}
 \nonumber
\end{split}\end{equation}
This susceptibility has maxima that are extended diagonally around ${\bf q}=(0,0,2\pi)$ in the $(q_x,0,q_z)$ plane as shown in Fig.~\ref{fig:CMC}\,(a). There are weak maxima, marked by red dots, which are shifted from  ${\bf q}=(0,0,2\pi)$ to   ${\bf q}=(0.603,0,0.986)\pi$ and ${\bf q}=(-0.603,0,3.014)\pi$.

{\it Classical Monte Carlo.-}
We perform classical Monte Carlo calculations for the Heisenberg Hamiltonian parameters given in Table~\ref{tab:exchangepars}. 
We perform the standard single spin-flip technique with the Metropolis updates.
The result is shown in Fig.~\ref{fig:CMC}\,(b)-(d). The specific heat shows two peaks at $T=21$\,K and at $T=10$\,K. This indicates that even though the two dominant exchange couplings $J_4$ and $J_5^\mathrm{A}$ are unfrustated, the frustrating coupling $J_2^\mathrm{C}$ leads to a significant reduction of the ordering temperature compared to the Curie-Weiss temperature of $\theta_{\rm CW}=-140$\,K. This is in good agreement with experiment where the material shows a pronounced ordering peak at $T=40.4$\,K. The peaks in the susceptibility (Fig.~\ref{fig:CMC}\,(b)) are less clearly separated. The type of ordering can be understood from Figs.~\ref{fig:CMC}~(c) and (d). Upon lowering the temperature, the dominant instability is at ${\bf q}=(0,0,2\pi)$ (Fig.~\ref{fig:CMC}~(c)). When $T$ is lowered further, the weak corrections due to
the frustration present in the Hamiltonian kick in, increasing the weight slightly away from ${\bf q}=(0,0,2\pi)$ (Fig.~\ref{fig:CMC}~(d)). Thus, the second ordering peak corresponds to the weak maxima marked by red dots in Fig.~\ref{fig:CMC}~(a).

{\it Conclusions.}
We have investigated the electronic structure and magnetic properties of the recently discovered \cuag compound, which combines strongly correlated Cu$^{2+}$ and Ag$^{2+}$ ions. This material bears many similarities with high-$T_\mathrm{c}$ cuprates, but also a number of remarkable differences, outlined in Table \ref{tab:0}. The differences stem from the fact that in this compound the sulfate ion SO$_4^{2-}$ plays the ligand role, as opposed to oxygen. As a result, the relevant hopping and exchange paths are longer-range, the antiferromagnetic ground state is highly unusual, and potential hole doping proceeds via pure O $p$ bands (rather than a hybridized Cu-O band, as in the cuprates). In addition, the ground state is altermagnetic, that is to say, sports spin-split Cu $d$ bands (which, however, as mentioned, are considerably removed from the Fermi level).

This collection of highly unusual properties make \cuag a fertile playground for exotic magnetism and superconductivity (under doping); while these are beyond the scope of the current paper, we hope that it will inspire further experimental and theoretical studies in this direction.

\acknowledgments
I.M. was supported by the Army Research
Office under Cooperative Agreement Number W911NF-22-2-0173.
M.S. was supported by Graduate School of Science, Kyoto University under Ginpu Fund and by JSPS KAKENHI grant No. 22H01181, No. 23K19032. Part of the computation in this work has been done using the facilities of the Supercomputer Center, the Institute for Solid State Physics, the University of Tokyo.
We acknowledge fruitful discussions with Ludovic Jaubert. We also thank Ginyat Khalliulin for critical reading of the manuscript.
Some of the images in the paper were created using VESTA software~\cite{VESTA}.

\bibliography{cuagso42}

\clearpage


\section*{Supplementary material}

# Highly unusual, doubly-strongly-correlated, altermagnetic, 3D analogue of parent compounds of high-$T_c$ cuprates
## – Supplementary Materials –

Harald O. Jeschke,[1] Makoto Shimizu,[2] and Igor I. Mazin[3]

[1]*Research Institute for Interdisciplinary Science, Okayama University, Okayama 700-8530, Japan*
[2]*Department of Physics, Graduate School of Science, Kyoto University, Kyoto 606-8502, Japan*
[3]*Department of Physics & Astronomy, George Mason University,
Fairfax, VA 22030, USA and Quantum Science and Engineering Center,
George Mason University, Fairfax, VA 22030, USA.*

(Dated: March 4, 2024)

## Additional DFT results

In Fig. S1, we identify the character of the bands of $CuAg(SO_4)_2$ near the Fermi level by showing the Ag 4$d$ and Cu 3$d$ orbital weights, respectively. The local coordinate system is appropriately chosen with $x$ and $y$ pointing towards nearest Ag-O or Cu-O bonds and $z$ pointing towards the apex of the elongated octahedra. The orbital character of the four bands at the Fermi level is dominated by Ag $4d_{x^2-y^2}$ and Cu $3d_{x^2-y^2}$.

Fig. S2 shows the integrated number of occupied and empty states for $CuAg(SO_4)_2$ in the lowest energy antiferromagnetic $\mathbf{q} = (0, 0, 2\pi)$ state. This is calculated with GGA+U at $U_{Ag} = 3.6$ eV, $U_{Cu} = 4.8$ eV and corresponds to the density of state plot in Fig. 3 of the main text. Note that for total electron numbers, the values have to be multiplied by 2 for spin and by multiplicity in the formula, *i.e.* 2 for S and 8 for O.

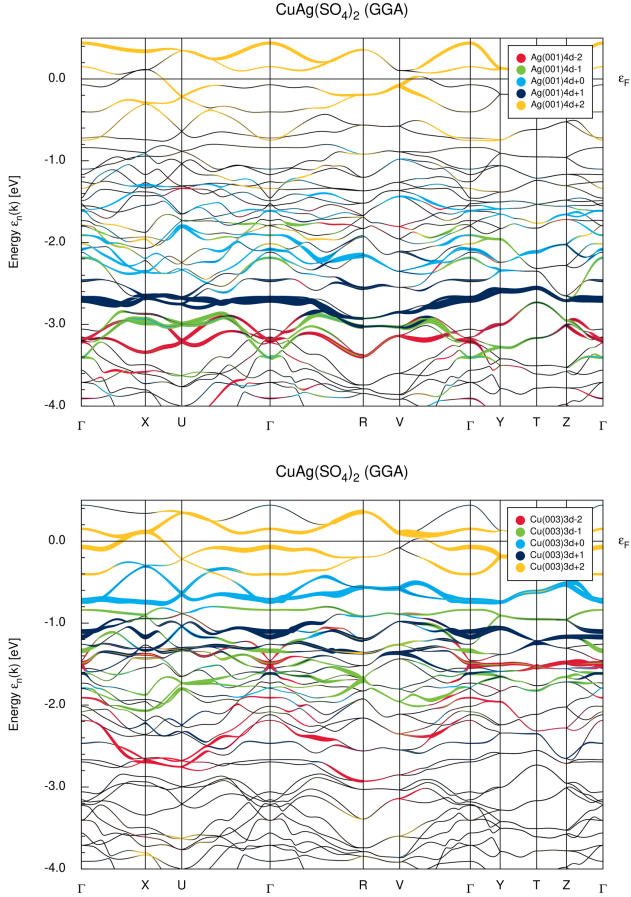

Figure S1. GGA band structure of $CuAg(SO_4)_2$ with orbital weights of Ag (top) and Cu (bottom). Ag 4$d$ and Cu 3$d$ orbital character, respectively, are highlighted: $d_{xy}$ ($m = -2$) red, $d_{yz}$ ($m = -1$) green, $d_{z^2}$ ($m = 0$) blue, $d_{xz}$ ($m = 1$) dark blue, $d_{x^2-y^2}$ ($m = 2$) yellow.

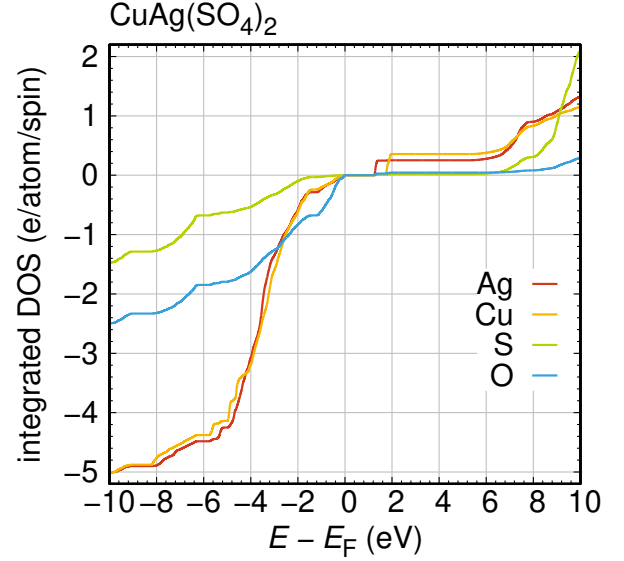

Figure S2. Integrated density of states of $CuAg(SO_4)_2$ at $U_{Ag} = 3.6$ eV, $U_{Cu} = 4.8$ eV in the lowest energy antiferromagnetic state. This corresponds to the density of states shown in Fig. 3 of the main text. The electron count is shown per atom and per spin.

## Comparison to cuprates

In Fig. S3, we show the density of states of the prototypical cuprate superconductor parent compound $La_2CuO_4$.

| $M1\,M2$ $d_{M1\,M2}$ (Å) | $U_{Ag}$ | $U_{Cu}$ | $J_1$ AgCu 3.579 | $J_2^A$ AgAg 4.734 | $J_2^C$ CuCu | $J_3$ AgCu 4.961 | $J_4$ AgCu 5.727 | $J_5^A$ AgAg 6.017 | $J_5^C$ CuCu | $\theta_{CW}$ |
|---|---|---|---|---|---|---|---|---|---|---|
| | 3.6 | 4.8 | -3.6(1.2) | 3.2(1.6) | 35.2(9) | 5.6(4) | 169.5(4) | 93.4(6) | -0.6(5) | -142 |
| | **3.65** | **4.87** | **-3.4(1.2)** | **3.0(1.6)** | **34.4(9)** | **5.4(4)** | **168.6(4)** | **91.9(6)** | **-0.5(5)** | **-140** |
| | 4.2 | 5.6 | -1.5(9) | 1.6(1.2) | 26.7(6) | 4.2(3) | 150.3(3) | 79.0(5) | -0.4(3) | -123 |
| | 4.8 | 6.4 | -0.2(6) | 0.6(9) | 20.1(5) | 3.4(2) | 134.2(2) | 68.5(4) | -0.3(3) | -108 |
| | 5.4 | 7.2 | 0.4(5) | 0.1(7) | 14.9(4) | 2.8(2) | 120.2(2) | 61.0(3) | -0.2(2) | -96 |

| $M1\,M2$ $d_{M1\,M2}$ (Å) | $U_{Ag}$ | $U_{Cu}$ | $J_6$ AgCu 6.134 | $J_7^A$ AgAg 6.215 | $J_7^C$ CuCu | $J_8^A$ AgAg 7.158 | $J_8^C$ CuCu | $J_{10}$ AgCu 8.332 | $J_{13}$ AgCu 8.846 | $J_{16}^A$ AgAg 9.266 | $J_{16}^C$ CuCu | $\theta_{CW}$ |
|---|---|---|---|---|---|---|---|---|---|---|---|---|
| | 3.6 | 4.8 | -2.6(7) | -0.4(7) | -3.9(5) | -2.2(3.6) | 0.5(1.2) | 0.5(4) | 0.4(2) | 1.0(4) | 0.0(3) | -142 |
| | **3.65** | **4.87** | **-2.5(7)** | **-0.4(7)** | **-3.8(5)** | **-2.1(3.6)** | **0.5(1.2)** | **0.5(4)** | **0.3(2)** | **0.9(4)** | **0.0(3)** | **-140** |
| | 4.2 | 5.6 | -2.0(5) | -0.4(5) | -2.7(3) | -1.6(2.6) | 0.3(9) | 0.4(3) | 0.2(2) | 0.7(3) | 0.0(2) | -123 |
| | 4.8 | 6.4 | -1.6(4) | -0.4(4) | -2.0(3) | -1.2(1.9) | 0.2(6) | 0.3(2) | 0.2(1) | 0.5(2) | 0.0(2) | -108 |
| | 5.4 | 7.2 | -1.2(3) | -0.3(3) | -1.5(2) | -0.9(1.4) | 0.2(5) | 0.2(2) | 0.1(1) | 0.4(2) | 0.0(1) | -96 |

Table S1. All calculated exchange parameters for CuAg(SO$_4$)$_2$. The line in bold face is interpolated by demanding that the set of couplings yield the experimentally observed Curie-Weiss temperature of $\theta_{CW} = -140$ K [1].

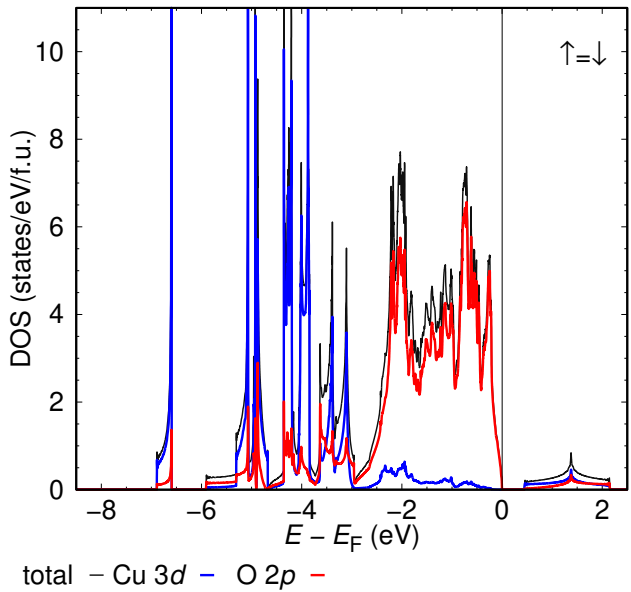

Figure S3. GGA+U density of states of La$_2$CuO$_4$ in the Néel state. Spin ↑ and spin ↓ are identical so only spin ↑ is shown.

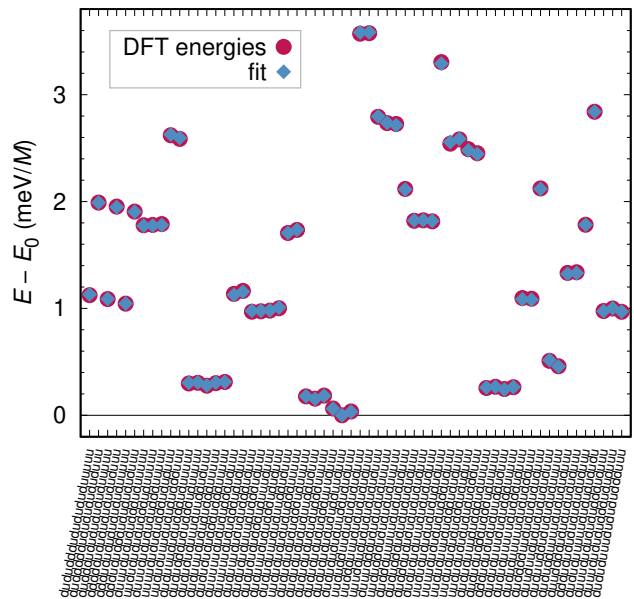

Figure S4. DFT energies for 60 spin configurations in a 5-fold unit cell of CuAg(SO$_4$)$_2$ at $U_{Ag} = 3.6$ eV, $U_{Cu} = 4.8$ eV, compared to the energies of the Heisenberg Hamiltonian with the fitted parameters given in the first line of Table S1. The fit is excellent.

### Additional DFT energy mapping details

In Table S1, we present the full results of the DFT energy mapping. They were obtained using a 5-fold unit cell of CuAg(SO$_4$)$_2$ containing ten formula units. This allows resolving all exchange interactions up to twice the nearest neighbor Ag-Cu distance of 3.58 Å. The Hund's rule coupling strength for Cu$^{2+}$ is fixed at $J_H = 1$ eV following previous work [2, 3]. We assume a factor 0.75 between $3d$ and $4d$ interaction values, thus fixing $J_H = 0.75$ eV for Ag$^{2+}$. The error bars given in brackets reflect the statistical errors of the fit for the energies of 60 distinct spin configurations.

We demonstrate that the DFT energy mapping approach works extremely well in CuAg(SO$_4$)$_2$ by showing, in Fig. S4, a comparison between DFT energies and energies of the fitted Heisenberg Hamiltonian.

### Details of the classical Monte Carlo method

In the classical Monte Carlo (CMC) calculations, we perform the standard single spin-flip technique with Metropolis updates. We apply as many Metropolis up-

dates as there are spins in the unit cell in each Monte Carlo step. For thermalization, we use 100,000 Monte Carlo steps. After thermalization, we use 10,000 Monte Carlo steps to measure physical quantities $A$. $A$ averages are calculated from spin configurations at intervals of 10 Monte Carlo steps. In this study, $A$ are total energy $E$, magnetization $M$ and spin structure factor

$$S(\mathbf{q}) = \sum_{ij} \mathbf{S}_i \cdot \mathbf{S}_j e^{i\mathbf{q}\cdot(\mathbf{r}_i-\mathbf{r}_j)} \tag{S1}$$

where $\mathbf{r}_i$ is the position of spin $i$. At the end of measurements, we calculate averages $\langle A \rangle_{\mathrm{MC}}$ of the measured physical quantities and calculate specific heat

$$C = k_{\mathrm{B}} \frac{\langle E^2 \rangle_{\mathrm{MC}} - \langle E \rangle_{\mathrm{MC}}^2}{T^2} \tag{S2}$$

and susceptibility

$$\chi = \frac{1}{N} \frac{\langle \mathbf{M}^2 \rangle_{\mathrm{MC}} - \langle |\mathbf{M}| \rangle_{\mathrm{MC}}^2}{k_{\mathrm{B}} T} \tag{S3}$$

where $k_{\mathrm{B}}$ is the Boltzmann constant, $T$ is temperature, $N$ is the number of spins in the unit cell, $\mathbf{M}$ is magnetization, and $E$ is total energy. We repeat this set of thermalization and measurements 160 times and calculate average and standard deviations of $C$, $\chi$ and $\langle S(\mathbf{q}) \rangle_{\mathrm{MC}}$ for all sets.

---



\end{document}